\documentclass[aps,prl,preprint]{revtex4-1}
\usepackage{graphicx}
\usepackage{amsfonts}
\usepackage{amsmath}
\usepackage{color}
\usepackage{bm}
\usepackage{amssymb}

\usepackage{multirow}
\usepackage{epstopdf}
\usepackage{natbib}

\newcommand{\beq}{\begin{equation}}
\newcommand{\eeq}{\end{equation}}

\def\R{\mathbf{\hat{R}}}

\def\R{\mathbf{\hat{R}}}
\def\mathbi#1{\textbf{\em #1}}

\def\bfrm#1{\textbf{\textrm #1}}

\def\colvecnext#1{
        #1
        \global\advance\colveccount-1
        \ifnum\colveccount>0
                \\
                \expandafter\colvecnext
        \else
                \end{pmatrix}
        \fi
}

\newcommand{\mat}[2][ccccccccccccccccccccccccccccccccccccccccccccccccccccccccccccc]{\left(
                                                \begin{array}{#1}
                                                    #2\\
                                                \end{array}
                                                \right) }

\begin{document}

\title{
Gaussian Process Model for Extrapolation of Scattering Observables for Complex Molecules: from Benzene to Benzonitrile}
\author{Jie Cui$^1$, Zhiying Li$^2$, Roman V. Krems$^1$}
\affiliation{$^1$Department of Chemistry, University of British Columbia, Vancouver, B.C., V6T 1Z1, Canada}
\affiliation{$^2$Department of Earth and Ocean Sciences, University of British Columbia, Vancouver, B.C., V6T 1Z1, Canada}

\date{\today}

\begin{abstract}

We consider a problem of extrapolating the collision properties of a large polyatomic molecule A-H to make predictions of the dynamical properties for another molecule related to A-H by the substitution of the H atom with a small molecular group X, without explicitly computing the potential energy surface for A-X. We assume that the effect of the $-$H $\rightarrow$ $-$X substitution is embodied in a multidimensional function with unknown parameters characterizing the change of the potential energy surface. 
We propose to apply the Gaussian Process model to determine the dependence of the dynamical observables on the unknown parameters. This can be used to produce an interval of the observable values that corresponds to physical variations of the potential parameters. 
   We show that the Gaussian Process model combined with classical trajectory calculations can be used to obtain the dependence of the cross sections for collisions of C$_6$H$_5$CN with He on the unknown parameters describing the interaction of the He atom with the CN fragment of the molecule. 
The unknown parameters are then varied within physically reasonable ranges to produce a prediction uncertainty of the cross sections. The results are normalized to the cross sections for He - C$_6$H$_6$ collisions obtained from quantum scattering calculations in order to provide a prediction interval of the thermally averaged cross sections for collisions of C$_6$H$_5$CN with He.

\end{abstract}

\maketitle

\section{Introduction}

This work is motivated by recent measurements of the thermalization dynamics of benzonitrile  C$_6$H$_5$CN in the cold gas of He \cite{bn-experiments}. The interpretation of the experiments requires the knowledge of the thermally averaged cross sections for collisions of C$_6$H$_5$CN with He. However, the potential energy surface for He -- C$_6$H$_5$CN interactions is, at present unknown and the accurate quantum dynamical calculation of the cross sections for He -- C$_6$H$_5$CN collisions is a formidable task. On the other hand, the interaction of a closely-related molecule C$_6$H$_6$ with He is well characterized and accurate quantum dynamical calculations of the cross sections for C$_6$H$_6$ -- He collisions have been previously reported \cite{benzene-he,benzene-he-note}. This raises the question: given the  cross sections for C$_6$H$_6$ -- He scattering is it possible to make predictions of the cross sections for 
C$_6$H$_5$CN -- He collisions? 

The experimentally measurable cross sections for elastic or inelastic collisions of molecules in the gas phase can be computed by means of quantum scattering or classical trajectory calculations \cite{scattering-theory}. Any such calculation uses as input the interaction potential energy surface (PES) that governs the molecular dynamics. In general, the PES must be computed for every collision system before the dynamical calculations.  The collision dynamics is typically sensitive to details of the PES and the collision properties of different molecules must be obtained from independent scattering calculations with the corresponding PESs. 
However, for large polyatomic molecules, the substitution of one atom or molecular group with another molecular group may alter only a small part of the global PES. In this case, instead of calculating the PES and the dynamical properties for the molecule with the substituted molecular group, 
one may consider to determine the effects of the substitution on the collision properties. 
This would allow one to make predictions about the collision properties of a specific molecule based on the known collision observables for another molecule.

Extrapolating the collision properties between different molecular systems is important for multiple applications. First of all, it can be used to significantly increase the range of molecules amenable to rigorous scattering theory analysis. Consider, for example, the collision properties of benzene C$_6$H$_6$ and benzene substitutes C$_6$H$_5$-X, where X is a halogen atom or a molecular group. 
The $D_{6h}$ symmetry of benzene reduces the numerical complexity of the quantum scattering calculations to a great extent \cite{benzene-he}. The absence of this symmetry in C$_6$H$_5$-X makes the scattering calculations much more computationally difficult, often impossible.  It is, therefore, desirable to develop an approach for predicting the scattering observables for benzene substitutes, given the computed scattering observables for C$_6$H$_6$. Second, it is often necessary to compare the collision properties of different molecules for calibrating the experimental observations and/or the design of new experiments \cite{experiments-polyatomic1,experiments-polyatomic2,experiments-polyatomic3}. However, the direct measurements for some molecular species may often be difficult or impossible. 

The extrapolation of the collision observables from one polyatomic molecule to another is a difficult task. Consider, for example, the interaction of C$_6$H$_6$ and C$_6$H$_5$-X with a He atom. 
The substitution of H by X distorts the highly symmetric PES for the He - molecule interaction. 
If X is a halogen atom, it may be possible to evaluate the effect of the substitution by 
computing the scattering cross sections for collisions of  C$_6$H$_5$X with He on a distorted benzene - He PES as functions of the distortion within a physically reasonable range of distortions. However, the problem becomes much more complex if $-$X is a molecular group, such as $-$CH$_3$ or $-$CN.
The introduction of the molecular group $-$X changes the number of the internal degrees of freedom and the distortion of the PES becomes a function of many parameters, determined by the relative positions of the atoms in the molecular group $-$X. For example, if $-$X is  $-$CN, the distortion of the  PES is a function of 6 parameters. It is impossible to analyze the effect of the $-$H $\rightarrow$ $-$CN substitution on the scattering observables by direct scattering calculations on a grid of these six parameters. 

To overcome this problem, we propose to apply the Gaussian Process (GP) model \cite{gp-model1,gp-model2,gp-model3}, used for machine learning applications in engineering technologies \cite{application-GP1,application-GP2,application-GP3}.  
Given a set of data depending simultaneously on multiple parameters, the GP model determines the correlations between the data points in the entire data sample and provides a highly efficient, non-parametric interpolation between the data points in the multi-dimensional space.  The Gaussian Process model is known to yield good prediction accuracy, when trained by 10 to 100 data points per dimension \cite{sample-size}.  An accurate six-dimensional hypersurface of scattering observables can thus be obtained with only 60 to 600 dynamical calculations at randomly chosen values of the 6 PES parameters. 
Here, we show that this can be exploited to provide a {\it prediction interval} of the scattering observables for a molecule A-X, given the scattering observable of the molecule A-H, where A and X are some molecular groups.  To provide predictions useful for the experiments in Ref. \cite{bn-experiments}, we use the GP model for the extrapolation of the scattering cross sections for benzene - He collisions to determine the prediction range of the scattering cross sections for benzonitrile - He collisions.

\section{Gaussian Process Model of Scattering Cross sections}

This work extends our previous contribution \cite{previous-work}, where we proposed to apply the GP model for the analysis of the sensitivity of the scattering dynamics of complex molecules to details of the PES. In the present work we assume that the scattering cross sections of a molecule A-H are known, either from an experiment or from a quantum dynamics calculation. The goal is to formulate the methodology to explore the effects of the substitution $-$H $\rightarrow$ $-$X on the scattering cross sections without explicitly computing the PES for A-X.  We begin with a general formulation and specialize the problem to the collision systems of benzene and benzonitrile with He in the next section.

We assume that X is a molecular group consisting of $N$ atoms. For simplicity, we consider the interaction of the molecules A-H and A-X with a structureless atom Rg and assume that the interaction of A-H and A-X with Rg is described by a single adiabatic PES. This implies that both of the molecules and the Rg atom have closed electronic shells. 
The substitution of a hydrogen atom with an $N$-atom molecular group adds $ 3 N-3$ degrees of freedom. The change of the PES due to this substitution depends on the coordinates specifying the position of each atom in the group $-$X relative to Rg. 
The PES for the interaction of the molecule A-X with Rg can be most generally written as 
\begin{eqnarray}
V_{\rm A-X} = V_{\rm A-H} + V_{\rm add}(\bm r, \bm R),
\label{pes}
\end{eqnarray}
where $V_{\rm A-H}$ is the PES for the interaction of the molecule A-H with Rg, 
the vector $\bm R = \bm R_{\rm A-X} - \bm R_{\rm Rg}$ specifies the separation between the position $\bm R_{\rm A-X}$ of the center of mass of the molecule A-X and the position $\bm R_{\rm Rg}$ of the Rg atom, and $\bm r = \left ( r_1, r_2,\cdots , r_{\cal N} \right)^{\top}$ is a vector with ${\cal N} = 3N$ components that represent the positions of the individual atoms in the molecular group $-$X relative to $\bm R_{\rm A-X}$. The allowed range of the coordinates $r_i$ is restricted by the parameters of the individual bonds in the molecular group $-$X.

   A rigorous calculation of the scattering cross sections for collisions of A-X with Rg requires the evaluation of the PES (\ref{pes}) as a function of $\bm r$ and $\bm R$, followed by the scattering calculations. 
 This is generally a prohibitively difficult task that we want to avoid. Instead, we propose to determine the {\it range} of collision cross sections given physically reasonable variations of the PES as a function of the coordinates $r_i$. The specific implementation of this procedure is illustrated in the next section.

     The coordinates $r_i$ can be written as $r_i = r_i^\circ + \delta r_i$, where $r_i^\circ$ are the coordinates of the atoms in the equilibrium geometry of the molecular group $-$X, and $\delta r_i$ describe the deviations from the equilibrium geometry.  The equilibrium geometry is generally unknown so the equilibrium coordinates $r_i^\circ$ must be treated as unknown parameters. 
      The value of the PES change $V_{\rm add}(\bm r, \bm R)$ at each point of $\bm r$ and $\bm R$ can be restricted on physical grounds to be within $\Delta V(\bm r, \bm R)$. 
For the methodology proposed here, it is necessary to represent $V_{\rm add}(\bm r, \bm R)$ by a series of analytical functions characterized by a finite set of parameters $\bm \alpha = (\alpha_1, ..., \alpha_k)^\top$, such that the variation of the individual parameters $\alpha_i$ within a certain range changes $V_{\rm add}(\bm r, \bm R)$ within $\Delta V(\bm r, \bm R)$. 
 The change of the PES in Eq. (\ref{pes}) thus becomes a function of these parameters:  $V_{\rm add}(\bm r, \bm R) \Rightarrow V_{\rm add}(\bm r, \bm R | \bm \alpha )$. 
 Since $r_i^\circ$ are assumed to be unknown, $3N$ of the parameters in the vector $\bm \alpha$ must correspond to $r_i^\circ$.


  With this formulation, the problem is reduced to determining the variation of the scattering cross sections as functions of $\alpha_1, ..., \alpha_k$. If the computation of the PES is to be avoided, one could -- in principle -- compute the scattering cross sections as functions of the individual parameters $\alpha_i$, thus producing a prediction interval of the scattering observables. 
    Assuming that $V_{\rm add}(\bm r, \bm R | \bm \alpha )$ can be parametrized by ten parameters $\alpha_i$ in addition to $3 N$ values of the equilibrium positions $r_i^\circ$, the total number of parameters is $3N + 10$. This ranges from 16 for a two-atom molecular group $-$X to 22 for a four-atom molecular group. It is clearly unfeasible to perform dynamical scattering calculations, whether quantum or classical, on a grid of these 22 parameters. We propose to use the GP model to determine the dependence of the cross sections on   $\bm \alpha$, in order to find the prediction interval of the cross sections corresponding to the range of the PES variations $\Delta V(\bm r, \bm R | \bm \alpha )$.

We consider the scattering cross section $\Omega$ as a function of $q$ parameters described by vector $\bm x$. The components of the vector $\bm x = \left ( x_1, x_2,\cdots , x_q \right)^{\top}$ are the collision energy, the internal energy of the molecule A-X and the parameters $\bm \alpha$.  The cross sections $ \Omega$ can be calculated by means of a classical trajectory method  \cite{ct-book} at fixed values of $(\alpha_1, ...., \alpha_k)$, fixed values of the collision energy and well-defined internal energies of the molecule. Given the calculated values of $\Omega$ at a small number of randomly chosen values of $(\alpha_1, ...., \alpha_k)$, we determine the correlations between different values of $\Omega$. These correlations are then used to train a GP model in order to make predictions of the cross sections for arbitrary values of $(\alpha_1, ...., \alpha_k)$ within a given interval. The number of computed cross sections  necessary to make accurate predictions can be estimated to be $10~\times$ the number of parameters \cite{sample-size}. 
With 22 parameters $\bm \alpha $ and the collision and internal energies as the additional free parameters, the number of $\Omega$ values required to determine an accurate dependence of $\Omega$ on $\bm \alpha $ should be about 240. Whether this number of calculations yields sufficient accuracy can be easily tested by computing the cross sections at new, arbitrary values of  $\bm \alpha$ and comparing the computed values with the GP model predictions.

 A GP can be viewed as a family of random functions normally distributed around a mean function. 
 We denote the GP by $F(\cdot)$.
 The realization $F(\bm x)$ of a GP at a particular site $\bm x$ of the multi-parameter space  is a value of a function randomly drawn from this family of functions and evaluated at $\bm x$. In the following, we will refer to $F(\cdot)$ as a GP or as a Gaussian random function.  A GP  is completely defined by its mean function $\mu(\cdot)$ and a covariance function $K(\cdot, \cdot)$. 
  We assume a GP with a constant variance $\sigma^2$ so that $K(\cdot, \cdot) = \sigma^2 R(\cdot, \cdot)$, where $R(\cdot, \cdot)$ is a correlation function. 
    The random outputs $F(\bm x)$ at fixed $\bm x$ thus form a normal distribution with mean $\mu(\bm x)$ and variance $\sigma^2$ and the multiple outputs at different $\bm x$ jointly follow a multivariate normal distribution \cite{adler1981geometry, cramer2013stationary}.

 The scattering cross section at each value of $\bm x$ is assumed to be a realization of a Gaussian process $F(\bm x)$. Given a set of cross sections computed at a finite number of values $\bm x$, the goal is to find $\mu(\cdot)$ and $R(\cdot, \cdot)$. 
 Since the correlation function is fitted to calculated data, the choice of $R(\cdot,\cdot)$ is somewhat flexible.
We use the following analytical form for the correlation function, known as the Mat\'ern correlation function  \cite{mitchell1990existence, cressie1993statistics, stein1999interpolation,abt1999estimating}:
\begin{eqnarray}
R(\mathbi{x},\mathbi{x}') =\prod_{i=1}^{q}\frac{1}{\Gamma(\nu)2^{\nu-1}}d_i^{\nu}\mathcal{K}_{\nu}(d_i)
\end{eqnarray}
where $d_i = \sqrt{2\nu}|x_i - x_i'|/\omega_i$, $\mathcal{K}_{\nu}(\cdot)$ is the modified Bessel function of order $\nu$ and $\omega_i$ are the unknown parameters representing the characteristic length scales of the correlation variations.
We fix the value of $\nu$ to $\nu=5/2$, which reduces the correlation function to
\begin{eqnarray}
R(\mathbi{x},\mathbi{x}') = \left\{ \prod_{i=1}^{q} \bigg(1+\frac{\sqrt{5}|x_i-x_i'|}{\omega_i}+\frac{5(x_i-x_i')^2}{3\omega_i^2}\bigg)\mathrm{exp}\bigg(-\frac{\sqrt{5}|x_i-x_i'|}{\omega_i}\bigg) \right\}. 
\end{eqnarray}
We find that the Mat\'ern correlation function yields more accurate results for the present application than the popular Gaussian correlation function \cite{mitchell1990existence, cressie1993statistics, stein1999interpolation,abt1999estimating} we used in the previous work \cite{previous-work}.
With the Gaussian correlation function, the GP is differentiable to any order. With the Mat\'ern correlation function with parameter $\nu$, the process is differentiable to order $k$ ($k<\nu$). Thus, with $\nu=5/2$, the GP is twice differentiable.


The mean of the Gaussian random function $F(\cdot)$ can be modelled as 
\begin{eqnarray}
\mu(\bm x) = \sum_{j=1}^{s} {h}_{j}(\bm{x})\beta_{j} = {\bf{h}}(\bm{x})^\top{\boldsymbol{\beta}}
\label{mean}
\end{eqnarray}
where ${\bf h} = \left ( h_1(\bm x), ..., h_s(\bm x) \right )^\top$ is a vector of $s$ regression functions \cite{gp-model2,gp-model3} and
$\boldsymbol{\beta} = (\beta_1,\beta_2,\cdots, \beta_s)^\top$ is a vector of unknown coefficients. 
If the dependence of the cross sections on any of the parameters in $\bm x$ is known, the regression functions $h_i(\bm x)$ can be chosen to mimic this dependence. This can make the GP model more efficient, i.e. fewer cross section values may be required to achieve the desired level of accuracy at arbitrary values of parameters. However, we note that the GP model does not rely on any specific form of the regression functions in Eq. (\ref{mean}). In the present work, we assume that $h_1 = 1$ and $h_{i>1} = 0$, which reduces Eq. (3) to a single unknown parameter $\beta$. The problem is thus reduced to finding the parameters ${\beta}$, $\sigma^2$ and $\boldsymbol{\omega}=(\omega_1,\omega_2,\cdots,\omega_q)^\top$ that provide the most accurate correlation function.


The GP model analysis begins with the computation of the cross sections at $n$ input vectors $\bm x_1, ..., \bm x_n$ randomly chosen to cover the allowed interval of the parameters.  These vectors are referred to as the training sites. The multiple outputs of a GP at the training sites  $\bm{Y}^n=\Big({F}(\bm{x}_1), 
{F}(\bm{x}_2),\cdots,{F}(\bm{x}_n)\Big)^{\top}$ follow a multivariate normal distribution
\begin{eqnarray}
\bm{Y}^n \thicksim \mathrm{MVN} (\mathbf{H}\boldsymbol{\beta},\sigma^2 \mathbf{A})
\end{eqnarray}
with the mean vector $\mathbf{H}\boldsymbol{\beta}$ and the covariance matrix 
$\sigma^2 \mathbf{A}$. Here, $\mathbf{H}$ is an $n \times s$ design matrix with $s$ regressors for each training site $\bm x_n$ as the matrix elements
\begin{eqnarray}
\mathbf{H} =  \left(\begin{array}{cccc} h_1(\bm{x}_1) & h_2(\bm{x}_1) & \cdots & h_s(\bm{x}_1) \\  h_1(\bm{x}_2) & \ddots & \ & \vdots \\ \vdots&  & \ddots &  \\h_1(\bm{x}_n) & \cdots &  & h_s(\bm{x}_n) \\\end{array} \right)
\end{eqnarray}
and $\mathbf{A}$ is a $n\times n$ matrix defined as
\begin{eqnarray}
\mathbf{A} =  \left(\begin{array}{cccc} 1 & R(\bm x_1, \bm x_2) & \cdots & R(\bm x_1, \bm x_n) \\  R(\bm x_2, \bm x_1) & 1& \ & \vdots \\ \vdots&  & \ddots &  \\R(\bm x_n, \bm x_1) & \cdots &  & 1 \\\end{array} \right)
\end{eqnarray}

If the coefficients $\boldsymbol{\omega}=(\omega_1,\omega_2,\cdots,\omega_q)^\top$
are known, the maximum likelihood estimators (MLE) of $\boldsymbol{\beta}$ and $\sigma^2$ are given in terms of $\bf{H}$ and $\bf{A}$ \cite{gp-model1,gp-model2}:
\begin{eqnarray}
\boldsymbol{\hat{\beta}} (\boldsymbol{\omega}) = (\mathbf{H}^{\top}\textbf{A}^{-1}\mathbf{H})^{-1}\mathbf{H}^{\top}\textbf{A}^{-1}\textbf{\textit{Y}}^n
\end{eqnarray}
\begin{eqnarray}
\hat{\sigma}^2 (\boldsymbol{\omega}) = \frac{1}{n}(\textbf{\textit{Y}}^n-\mathbf{H}\boldsymbol{\beta})^{\top}{\bf{A}}^{-1}(\textbf{\textit{Y}}^n-\mathbf{H}\boldsymbol{\beta}),
\end{eqnarray} 
where the hat over the symbol denotes the MLE. 
To find the MLE of $\boldsymbol{\omega}$, we maximize the log-likelihood function
\begin{eqnarray}
\textrm{log}\mathcal{L}(\boldsymbol{\omega}|\bm{Y}^n) = -\frac{1}{2}\left[n\textrm{log}\hat{\sigma}^2+ \textrm{log}(\textrm{det}(\textbf{A})) +n \right]
\end{eqnarray}
numerically by an iterative computation of the determinant  $|\textbf{A}|$ and the matrix inverse $\textbf{A}^{-1}$.

The goal is to make a prediction of the cross section at an arbitrary input vector $\bm x = \bm x_0$, given the values of the cross sections at the training sites. 
The values $Y_0=F(\bm{x}_0)$ obtained by multiple realizations of the GP at $\bm x_0$
and the multiple outputs of the GP at training sites 
$\bm{Y}^n=\Big({F}(\bm{x}_1), {F}(\bm{x}_2),\cdots,{F}(\bm{x}_n)\Big)^{\top}$ 
are jointly distributed as
\begin{eqnarray}
\mat{Y_0\\ \bm{Y}^n}  \thicksim \mathrm{MVN} \left\{\mat{\textbf{\textrm{h}}(\bm{x}_0)^{\top} \\ \textbf{H}}\boldsymbol{\beta}
, \sigma^2 \mat{1 & \mathbf{A}_0^{\top}\\ \mathbf{A}_0&\mathbf{A} }\right\}
\end{eqnarray}
where 
 $\mathbf{A}_0=(R(\bm{x}_0,\bm{x}_1),R(\bm{x}_0,\bm{x}_2),\cdots,R(\bm{x}_0,\bm{x}_n))^{\top}$ is a column vector specified by the correlation function $R(\cdot|\boldsymbol{\hat \omega})$ with the MLE of $\boldsymbol{\omega}$. 
This means that 
the conditional distribution of possible values $Y_0=F(\bm{x}_0)$ given the values $\bm{Y}^n$ is a normal distribution 
\begin{eqnarray}
Y_0|\bm{Y}^n, \boldsymbol{\beta}, \sigma^2, \boldsymbol{\omega}\thicksim \mathrm{N} (m(\bm{x}_0)^*,\sigma^{*2}_z(\bm{x}_0))
\end{eqnarray}
with the conditional mean and variance given by
\begin{eqnarray}
\label{mean-prediction}
m(\bm{x}_0)^*&=& \bfrm{h}(\bm{x}_0)^{\top}\boldsymbol{\beta}+\mathbf{A}_0^{\top}\mathbf{A}^{-1}(\bm{Y}^n -\textbf{H}\boldsymbol{\beta} ) \\
\sigma^{*2}(\bm{x}_0) &=& \sigma^2 (1- \mathbf{A}_0^{\top}\mathbf{A}^{-1}\mathbf{A}_0).
\end{eqnarray}
Note that the conditional mean given by Eq. (\ref{mean-prediction})  is shifted from the unconditional mean equal, in our case, to $\beta$ due to the point-to-point correlations $R(\cdot, \cdot)$. 
Replacing $\bm{Y}^n$ in Eq. (\ref{mean-prediction}) with the vector of the known cross section values ${\bm \Omega} = \Big(\Omega(\bm x_1), \cdots, \Omega(\bm x_n)  \Big)^\top$, we obtain the GP model prediction for the value of the cross section at $\bm x_0$
\begin{eqnarray}
\label{mean-prediction-sigma}
\Omega(\bm{x}_0)&=& \bfrm{h}(\bm{x}_0)^{\top}\boldsymbol{\beta}+\mathbf{A}_0^{\top}\mathbf{A}^{-1}(\bm{\Omega} -\textbf{H}\boldsymbol{\beta} ).
\end{eqnarray}

\section{From Benzene to Benzonitrile}

In this section we apply the GP model 
to predict the cross sections for elastic scattering of benzonitrile (C$_6$H$_6$CN) with He, required for the interpretation of the experiments on cooling polyatomic molecules in a buffer gas of He \cite{bn-experiments}.
The PES for C$_6$H$_5$CN - He is currently unknown. The quantum scattering calculations of cross sections for collisions of C$_6$H$_5$CN with He are prohibitively difficult. At the same time, the interaction of benzene (C$_6$H$_6$) with He can be accurately described using a semi-empirical bond-additive method \cite{pirani2001atom,pirani2004atom} and the cross sections for collisions of C$_6$H$_6$ with He can be computed using an accurate coupled states approach \cite{cs-polyatomic1,cs-polyatomic2} based on the time-independent quantum scattering theory \cite{benzene-he}. In the present section, we show how the GP model can be used to obtain the range of the cross sections for He - C$_6$H$_5$CN collisions, 
given the cross sections for He - C$_6$H$_6$ collisions.



We use the semi-empirical approach of Pirani and coworkers \cite{pirani2001atom,pirani2004atom} for constructing the PES for the interactions of the polyatomic molecules with He. 
 This method treats a polyatomic hydrocarbon molecule as an ensemble
of C-C and C-H bonds and represents the PES for the molecule - He interaction as a sum of pairwise He - CH bond and He - CC bond interaction energies, optimized based on
 accurate {\it ab initio} calculations and measurements of bond polarizabilities.  This method of representing the PES is particularly well suited for the analysis of the $-H$ $\rightarrow$ $-X$ substitution on the molecular scattering properties using the GP model described in the previous section.

Within the approach of Pirani and coworkers \cite{pirani2004atom},  each \textrm{He-CH} and \textrm{He-CC} bond interaction
 is represented by the following analytical function: 
 \begin{eqnarray}
V_{ab}(r,\theta)=\epsilon(\theta)\Bigg [\bigg(\frac{3}{2+2x^2}\bigg)\cdot\bigg(\frac{1}{x}\bigg)^{10+4x^2} - \bigg(\frac{5+2x^2}{2+2x^2}\bigg)\cdot \bigg(\frac{1}{x}\bigg)^{6} \Bigg]
\end{eqnarray}
where
\begin{eqnarray}
\epsilon(\theta)&=&\epsilon_{\perp} \sin^2(\theta)+\epsilon_{\parallel}\cos^2(\theta),\\
r_m(\theta)&=&r_{m\perp}\sin^2(\theta)+r_{m\parallel}\cos^2(\theta),
\end{eqnarray}
$r$ is the distance between the atom and the center of the bond, and $x$ is the reduced distance $x={r}/{r_m(\theta)}$, where $r_m(\theta)$ is the position of the potential well and $\epsilon(\theta)$ is the energy at the bottom of the potential well. 
The parameters $\epsilon_{\perp} ,\epsilon_{\parallel},r_{m\perp}$ and $r_{m\parallel}$ represent the well depth and location for the parallel and perpendicular approaches of the He atom to the corresponding bond and $\theta$ is the angle 
between the bond axis and the vector connecting the He atom to the center of the bond axis.
The eight  parameters specifying the PES for the He - C$_6$H$_6$ interaction 
are given in Table \ref{table51}.

\begin{table}[h!]
\centering
\begin{tabular}{ ccccc c}
\hline
System&$r_{m\perp}$&$r_{m\parallel}$&$\epsilon_{\perp}$&$\epsilon_{\parallel}$ & Reference\\
\hline
Aromatic CC-He& 3.583& 4.005 & 0.860& 0.881 & \cite{pirani2004atom} \\ 
Non-aromatic CC-He& 3.09& 4.10 & 1.03& 0.66 & \cite{private-communication}\\ 
CH-He& 3.234& 3.480& 1.364& 1.016 & \cite{pirani2004atom}\\ 
\hline
\end{tabular}
\caption{Parameters for the \textrm{He-CH}, \textrm{He-CC} aromatic and \textrm{He-CC} non-aromatic  bond interactions. The distances are given  in \AA $\,$ and the energies are in meV.}  
\label{table51}
\end{table}

In order to construct the PES for the He - C$_6$H$_5$CN interaction, we use the same approach. However, the PES for the He - C$_6$H$_5$CN system must include, in addition to the aromatic CC - He and CH - He interactions, the parameters describing the interaction of the He atom with the non-aromatic single CC bond and the $-$CN bond fragment. 
The parameters for the He - non-aromatic, single CC bond interaction are known from the work in Refs. \cite{pirani2004atom,private-communication} and are listed in Table I. 
The parameters for the He - CN bond interaction are, at present, unknown and we propose to treat them as variable parameters. The PES for He - C$_6$H$_5$CN collisions is thus parametrized by twelve parameters listed in Table  \ref{table51} and four unknown parameters $r_{m\perp}^\ast$, $r_{m\parallel}^\ast$, $\epsilon_{m\perp}^\ast$ and $\epsilon_{m\parallel}^\ast$ characterizing the locations and energies of the well depths arising from the perpendicular and parallel approaches of the He atom to the $-$CN bond fragment in the molecule. We treat these parameters as variables in the input vector $\bm x$ of the GP model. The vector $\bm x$ is thus a five-dimensional (5D) vector, containing four parameters $r_{m\perp}^\ast$, $r_{m\parallel}^\ast$, $\epsilon_{m\perp}^\ast$ and $\epsilon_{m\parallel}^\ast$ and the collision energy. In this work, we fix the internal energy of the molecule to correspond to the ground rotational state.

Given that the size of the N atom is smaller than the size of the C atom and that the C-N bond is more polarized than the C-C bond, one should expect that the He - CN interaction parameters should lead to a larger potential depth but smaller equilibrium distance than the parameters of the He - CC bond interactions. This puts the upper limit on the values of $r_{m\perp}^\ast$ and $r_{m\parallel}^\ast$ and the lower limit on the values of $\epsilon_{m\perp}^\ast$ and $\epsilon_{m\parallel}^\ast$. In order to explore the dependence of the collision cross sections on the variation of these parameters, we construct 
100 different PESs for the He - C$_6$H$_5$CN system with the variable parameters in the following ranges: $r_{m\perp}^\ast \thicksim \rm{Unif}  (2.5,3.6)$ \AA, $r^\ast_{m\|} \thicksim \rm{Unif}  (3.0,4.0)$ \AA, $\epsilon^\ast_{\perp} \thicksim \rm{Unif}  (0.86,2.5)$ meV, $\epsilon^\ast_{\|} \thicksim \rm{Unif} (0.88,2.5)$ meV.

The substitution of the H atom in benzene with the molecular group CN is expected to introduce a local change of the global PES. To illustrate this, we plot in Figure 1 the cross sections of the PES for the He - C$_6$H$_5$CN and He - benzene interactions for various geometries of the atom - molecule approach. 
Figure 1 illustrates that the CN molecular group leads to a significant change of the potential energy only when the He atom approaches the CN side of the molecule. This justifies the representation (\ref{pes}) of the global PES and the approach adopted here.



\section{Results}

Since the quantum scattering calculations of cross sections for He - C$_6$H$_5$CN collisions are prohibitively difficult, we use the classical trajectory method developed in Refs. \cite{zhiying1, benzene-he, jie-zhiying} for the dynamical calculations for benzonitrile. 
Using the classical trajectory computations, as described in detail in Ref. \cite{zhiying1}, we compute the scattering cross sections for 100 combinations of different PESs and collision energies. 
Each computation represents an average of 5000 trajectories. The cross sections are computed using the corrected version of Eq. (23) in Ref. \cite{benzene-he, benzene-he-note}.
These cross sections are then used to train the GP model in order to provide the global dependence of the cross sections on the collision energy and the four unknown interaction potential parameters.  The resulting ranges of the cross sections are then scaled by the ratio $ \tilde \Omega_{\rm qm}/ \tilde  \Omega_{\rm ct}$, where $\tilde  \Omega_{\rm qm}$ is the thermally averaged cross section for He - benzene collisions computed with the coupled states method \cite{cs-polyatomic1,cs-polyatomic2} and $ \tilde \Omega_{\rm ct}$ is the same cross section computed with the classical trajectory method used for benzonitrile.

Our first goal is to obtain the range of the cross sections corresponding to a physical variation of the He - CN bond interaction parameters $r_{m\perp}^\ast$, $r^\ast_{m\|}$, $\epsilon^\ast_{\perp}$, $\epsilon^\ast_{\|}$ and the atom - molecule collision energy.
Figure 2 shows the variation of the cross sections as a function of one of the PES parameters, with the remaining PES parameters and the collision energy chosen for each point at random. As evident from Figure 2, the scattered points cannot be assumed to follow any well-defined dependence.



  We obtain the global dependence of the cross sections on the four PES parameters and the collision energy by computing the cross sections at 100 points, placed randomly and quasi-uniformly, in this 5D parameter space. These points are used to train the GP model as discussed in the previous section. 
Given the GP model, we compute the global 5D surface illustrated in Figure 3. In order to demonstrate the accuracy of the global 5D surface, we computed the cross sections at a different set of 100 points randomly chosen in the 5D parameter space. Figure 4 compares the predicted values with the computed values for these 100 points. 

In order to quantify the error of the GP model, we compute the empirical root mean squared error (ERMSE) defined as 
 \begin{eqnarray}
 \varepsilon_E = \sqrt{\frac{1}{n}\sum_{i=1}^{n}(y_i-\hat{y}_i)^2}
\label{error}
  \end{eqnarray}
and the scaled root mean squared error (SRMSE) defined by 
 $\varepsilon_S=\varepsilon_E/{(y_{\rm max}-y_{\rm min})}$.
 In Eq. (\ref{error}), $n$ is the number of the cross section values, $y_i$ represents the values of the computed cross sections and $\hat{y}_i$ -- the value predicted by the GP model. 
 For the model with only 50 scattering calculations used as training points, 
$\varepsilon_E = {{2.17}}$ \AA$^2$ and $\varepsilon_S = {{4.4}}~\%$. If the number of the scattering calculations is increased to {{100}}, the errors decrease to $\varepsilon_E = {{1.77}}$ \AA$^2$ and $\varepsilon_S = {{3.6}}~\%$.

Given the GP model of the collision cross sections trained by the computations at random values of 
$r_{m\perp}^\ast$, $r^\ast_{m\|}$, $\epsilon^\ast_{\perp}$, $\epsilon^\ast_{\|}$ and the atom - molecule collision energy, we can use Eq. (\ref{mean-prediction-sigma}) to examine the variation of the collision energy dependence of the cross sections as the four parameters are varied within the physical ranges specified in the previous section.  This produces a band of the cross section values for each value of the collision energy presented in Figure 5. The results illustrate that the physical variation of the PES parameters describing the interaction of the He atom with the CN group changes the cross sections by less than $33$ \%, with the percentage defined as the difference between the maximum and minimum values divided by the mean value. The computation of the cross sections for benzonitrile - He collisions with an accurate PES must fall within the grey area of Figure 5 with the probability {{95}} \%. 

Since the GP model provides the global dependence of the cross sections on the underlying parameters, it is possible to perform the analysis of the sensitivity of the cross section variations to the individual PES parameters by using the functional analysis of variance decomposition \cite{saltelli2009sensitivity, saltelli2008global, roustant2012dicekriging}. The results shown in Figure 6 illustrate that, of the four unknown parameters, $\epsilon^\ast_{\|}$ has the biggest effect on the variation of the cross sections. Figure 6 also shows that the effect of the individual parameters is largely  {{correlated with}} the values of the other parameters. This means that the prediction interval in Figure 5 must be obtained by the simultaneous variation of each of the underlying parameters and cannot be obtained by a simple variation of one of the parameters with the other parameters kept fixed at a few random values. 

The final results of this work are presented in Figure 7, showing the thermally averaged cross sections for benzene - He collisions computed with the quantum CS approach ({{blue}} squares) and the interval of the thermally averaged cross sections for benzonitrile - He collisions obtained by integrating the results in Figure 5 with the Maxwell-Boltzmann distribution of collision energies and scaling the thermally averaged cross sections by the ratio $ \tilde \Omega_{\rm qm}/ \tilde  \Omega_{\rm ct}$. Figure 7 shows that the presence of the molecular group CN enhances the elastic scattering cross sections by a factor of {{1.1 -- 1.5}}. The uncertainty of the parameters in the He - CN group interaction leads to the uncertainty  $\sim 27 \%$ in the final thermally averaged cross section.  This percentage is defined as the difference between the maximum and minimum values of the grey band in Figure 7 divided by the corresponding mean value.

\section{Summary}

In the present work we consider a general problem of extrapolating the known collision properties of a complex polyatomic molecule A-H to make predictions for another molecule related to A-H by the substitution of the H atom with a small molecular group.  Using the example of C$_6$H$_6$ -- He and 
C$_6$H$_5$CN -- He collision systems, we show that the $-$H $\rightarrow$ $-$CN substitution leads to a local modification of the global potential energy surface. 
While the quantitative effect of the $-$H $\rightarrow$ $-$CN substitution on the PES is unknown, it can be parametrized by a finite number ($k$) of parameters $(\alpha_1, ...., \alpha_k)$, collectively denoted by $\bm \alpha$.  With this parametrization, the effect of the $-$H $\rightarrow$ $-$CN substitution on the collision observable $\Omega$ is embodied in a multidimensional function $\Omega(\bm \alpha)$. Even if the parameters $\bm \alpha$ are unknown, once the function of $\Omega$ on $\bm \alpha$ is determined, it can be used to obtain the prediction interval for $\Omega$ given physically reasonable variation of $\bm \alpha$. 

  We propose to apply the Gaussian Process model to determine $\Omega (\bm \alpha)$. The model requires about 10 $\times$ ($k+1$) calculations of the scattering observables to provide accurate results. Thus, with $k=4$, an accurate five-dimensional dependence of $\Omega$ on $\bm \alpha$ and on the collision energy can be obtained with about 50 dynamical calculations. We showed that this procedure can be used to obtain an accurate dependence of the cross sections for collisions of C$_6$H$_5$CN with He on the parameters describing the interaction of the He atom with the CN fragment of the molecule. The results are then compared with the cross section for He - C$_6$H$_6$ collisions known from the quantum scattering calculations in order to provide a prediction interval of the thermally averaged cross sections for collisions of C$_6$H$_5$CN with He. This allowed us to obtain the prediction interval (Figure 7) of the collision cross sections for C$_6$H$_5$CN -- He collisions without the knowledge of the potential energy surface. 

This work illustrates that the Gaussian Process model can be used for a variety of applications in molecular dynamics research. For example, once the function $\Omega(\bm \alpha)$ is obtained, it can be used to perform the sensitivity analysis (such as in Figure 6) to determine which of the PES parameters is more or less important in determining the collision observable $\Omega$. 
The dependence of $\Omega$ on the collision energy, which is treated as one of the unknown model parameters, can be used for an efficient integration of the collision properties to obtain thermally averaged observables. The function $\Omega(\bm \alpha)$ can be used to integrate out the dependence of the collision observables on the PES parameters, thereby minimizing the uncertainties associated with inaccuracies of the PES calculations and providing the error bars associated with the uncertainties of the PES calculations. 

The computational effort associated with training the Gaussian Process model is determined by matrix inversion and scales as the third power of the number of training points. Given that the number of training points required for accurate predictions is typically 10 times the number of unknown parameters and that it has now become routine to invert matrices with the dimension of 1000 $\times$ 1000, the methodology proposed here can be applied to determine the dependence of the scattering observables on up to 100 unknown parameters. It is thus easy to envision an application where the Gaussian Process model is used to characterize the dependence of a dynamical observable on all PES parameters describing the interaction of two polyatomic molecules. This dependence can be used to determine which of the atoms in the two molecules are important for the detectable outcome of the molecule - molecule interaction and which of the atoms can be parametrized by simple functions irrelevant for the outcome of the interaction. 

\clearpage
\newpage

\section{Acknowledgment}

We thank Dr. Massimiliano Bartolomei for sending us the parameters of the interaction potentials ans for allowing us to include his unpublished data in Table I. 
This work is supported by NSERC of Canada. 

\clearpage
\newpage

\begin{figure}[ht]
\label{figure1}
\begin{center}
\includegraphics[scale=0.5]{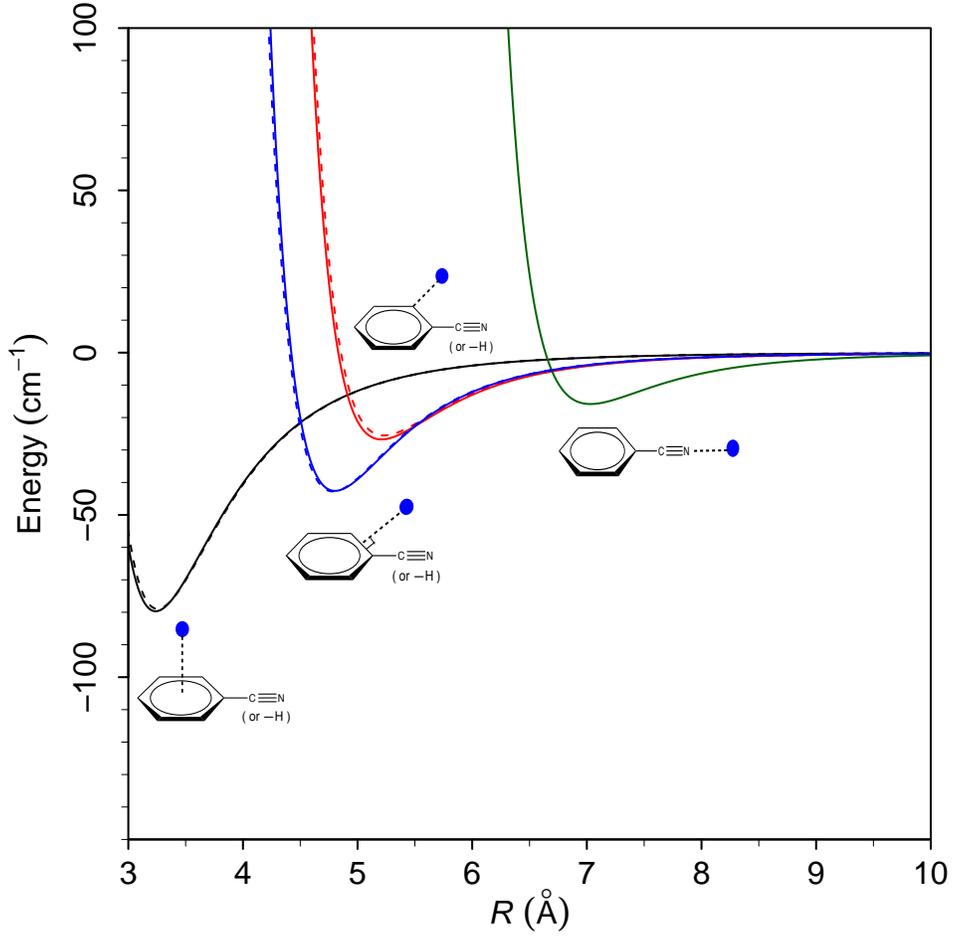}
\end{center}
\caption{Radial dependence of the interaction energy of benzonitrile with He averaged over four variable parameters $r_{m\perp}^\ast$, $r_{m\parallel}^\ast$, $\epsilon_{m\perp}^\ast$ and $\epsilon_{m\parallel}^\ast$ in the out-of-plane configuration (black solid line), the H-vertex-in-plane configuration (red solid line), the CN-vertex-in-plane configuration (solid green line) and the side-in-plane configuration (blue solid line) compared with the corresponding interaction energy of benzene with He (black dashed line for the out-of-plane configuration, red dashed line for the vertex-in-plane configuration, and blue dashed line for the side-in-plane configuration). The relative geometries of the four limiting configurations are depicted near each curve. 
}
\end{figure}

\begin{figure}[ht]
\label{figure2}
\begin{center}
\includegraphics[scale=0.4]{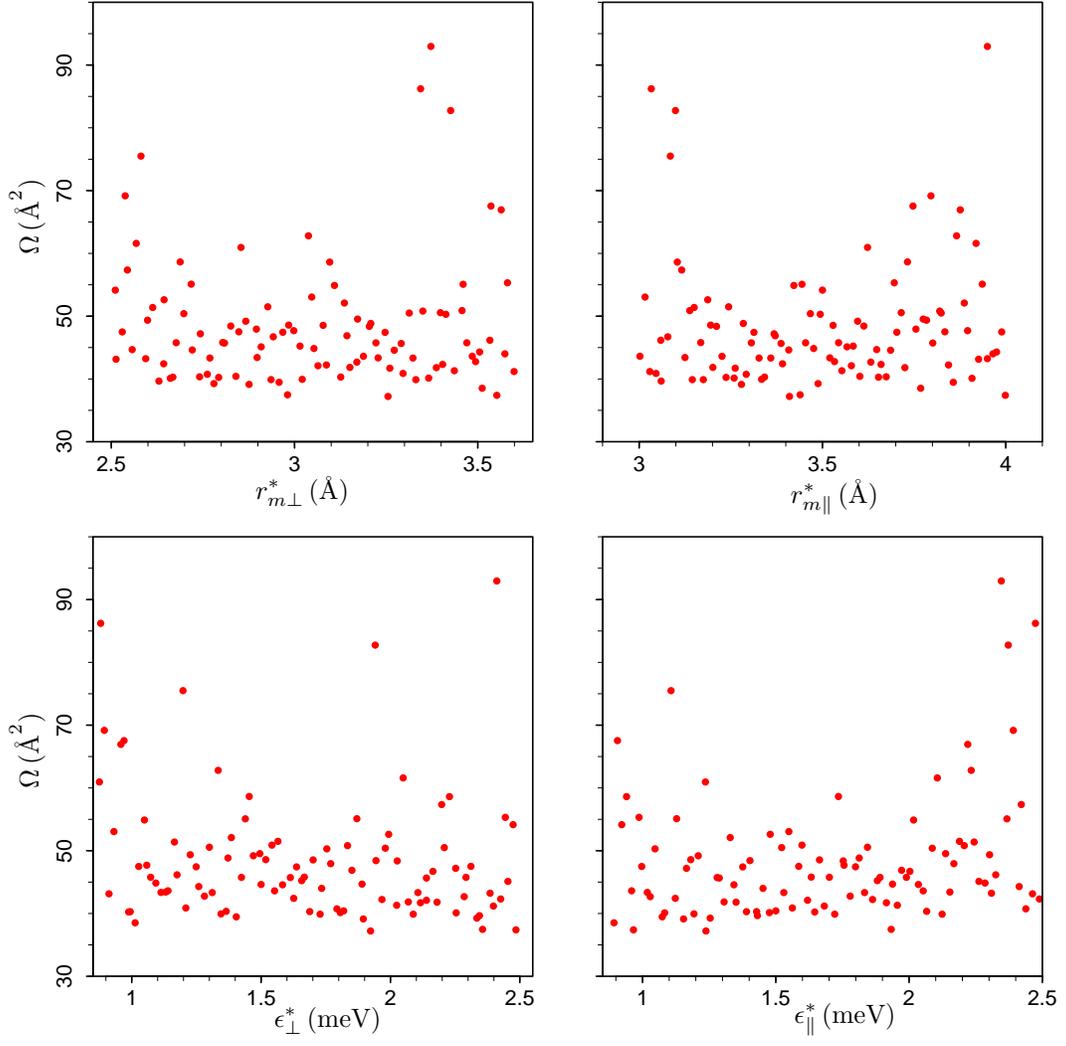}
\end{center}
\caption{The dependence of the elastic scattering cross section on each of the four PES parameters for benzonitrile-He collisions. The other three parameters and the collision energy are chosen at random for each point shown. 
}
\end{figure}

\begin{figure}[ht]
\label{figure3}
\begin{center}
\includegraphics[scale=0.5]{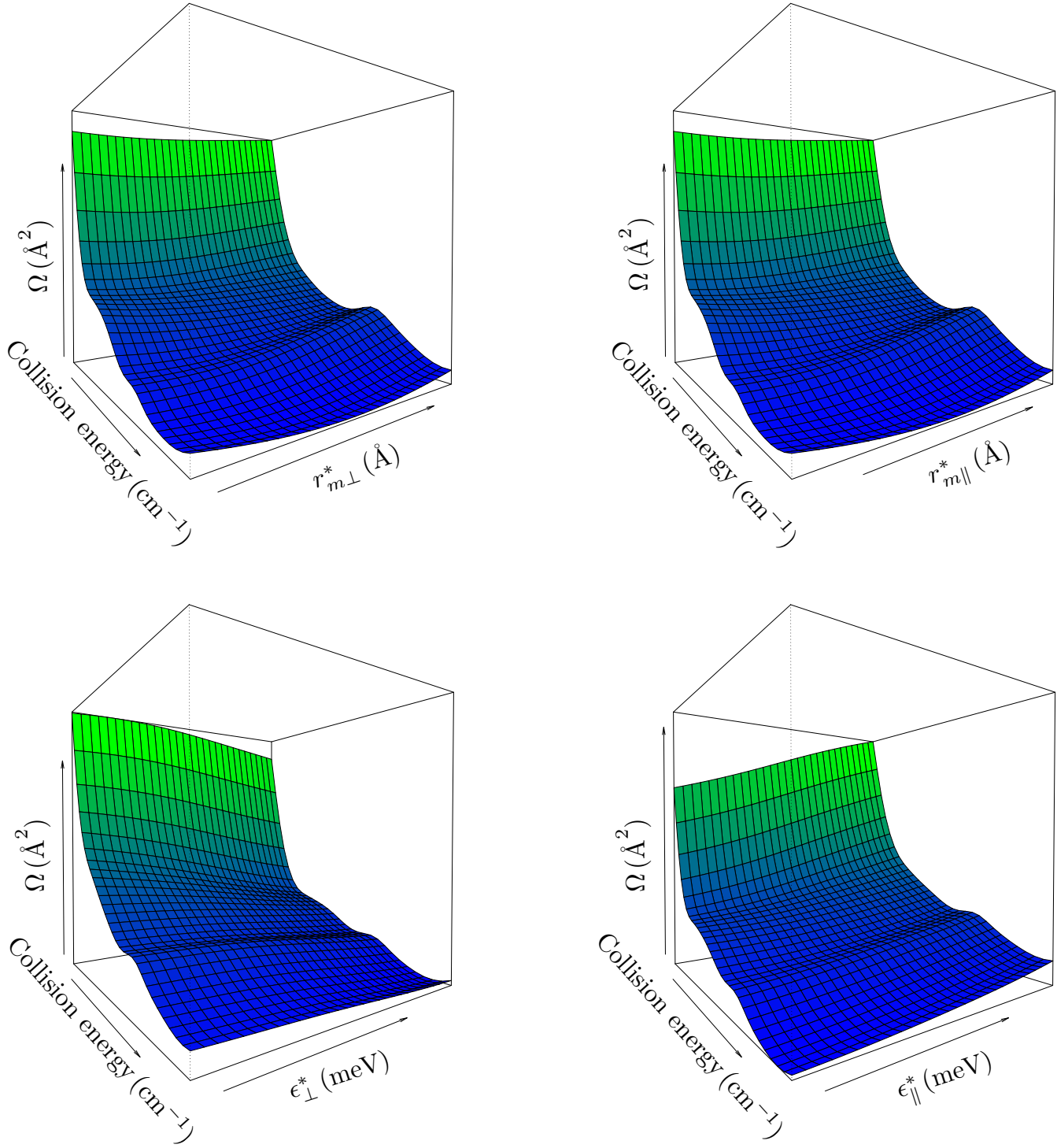}
\end{center}
\caption{The global 5D response surface produced by the GP model illustrated in three dimensions, with the collision energy and one of the PES parameter plotted on the x- and y-axises. The cross sections shown are averaged over the other three PES parameters.
}
\end{figure}

\begin{figure}[ht]
\label{figure4}
\begin{center}
\includegraphics[scale=0.8]{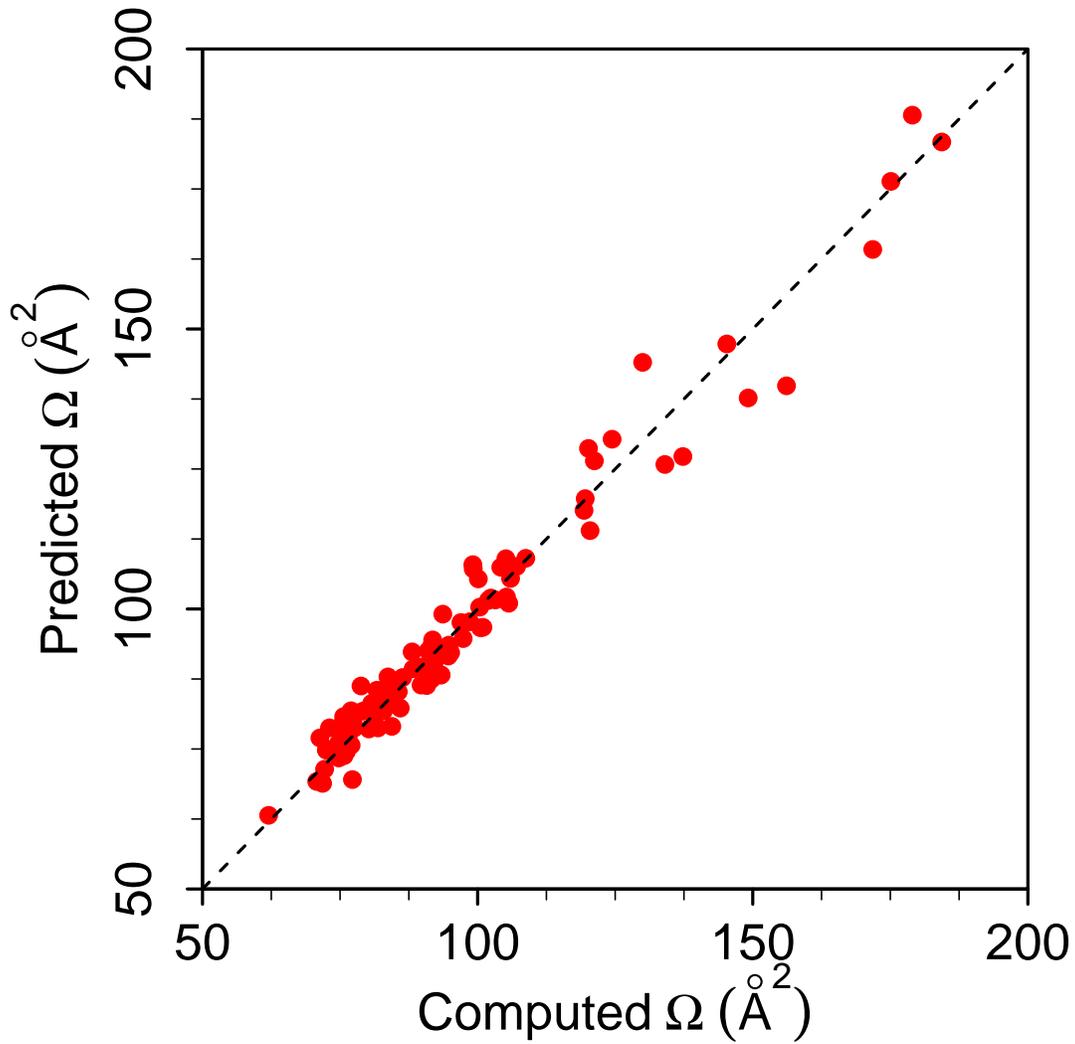}
\end{center}
\caption{Accuracy of the GP model with variable PES parameters for the prediction of the elastic scattering cross sections. The scatter plot compares the predicted values with the computed values at 100 points. The error of the GP model is the deviation of the points from the diagonal line.
}
\end{figure}

\begin{figure}[ht]
\label{figure5}
\begin{center}
\includegraphics[scale=0.5]{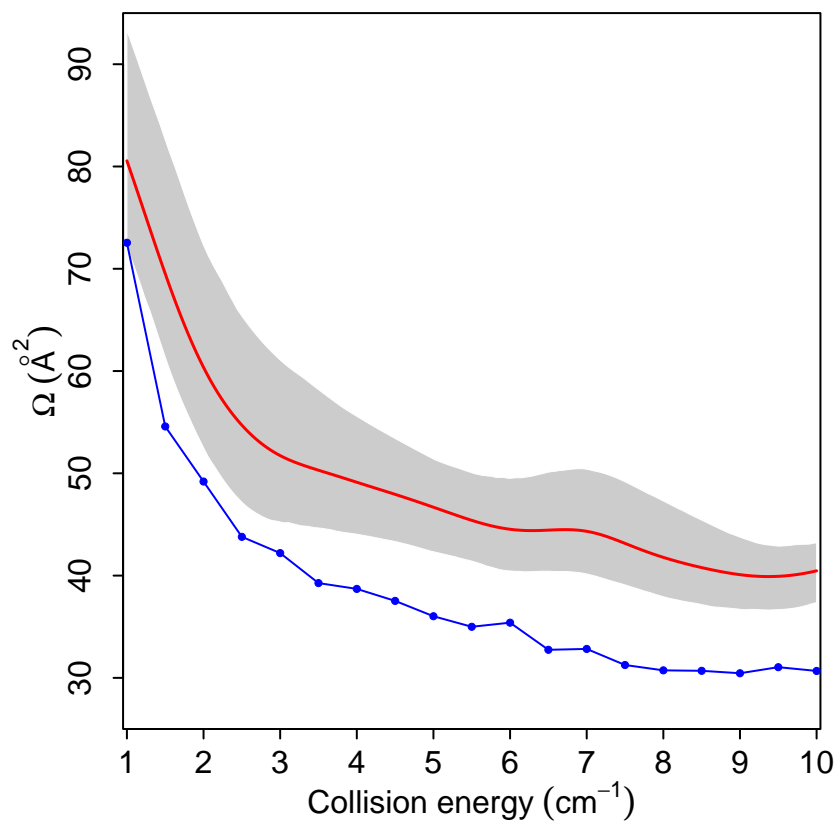}
\end{center}
\caption{Collision energy dependence of the elastic scattering cross section averaged over four PES parameters $r_{m\perp}^\ast$, $r_{m\parallel}^\ast$, $\epsilon_{m\perp}^\ast$ and $\epsilon_{m\parallel}^\ast$ for benzonitrile-He collisions (red solid line) and the associated 95$\%$ prediction interval (grey band). The line with blue circles shows the corresponding cross section for benzene-He collisions computed by the classical trajectory method. The results are for collisions of molecules in the ground rotational state. 
}
\end{figure}

\begin{figure}[ht]
\label{figure6}
\begin{center}
\includegraphics[scale=0.5]{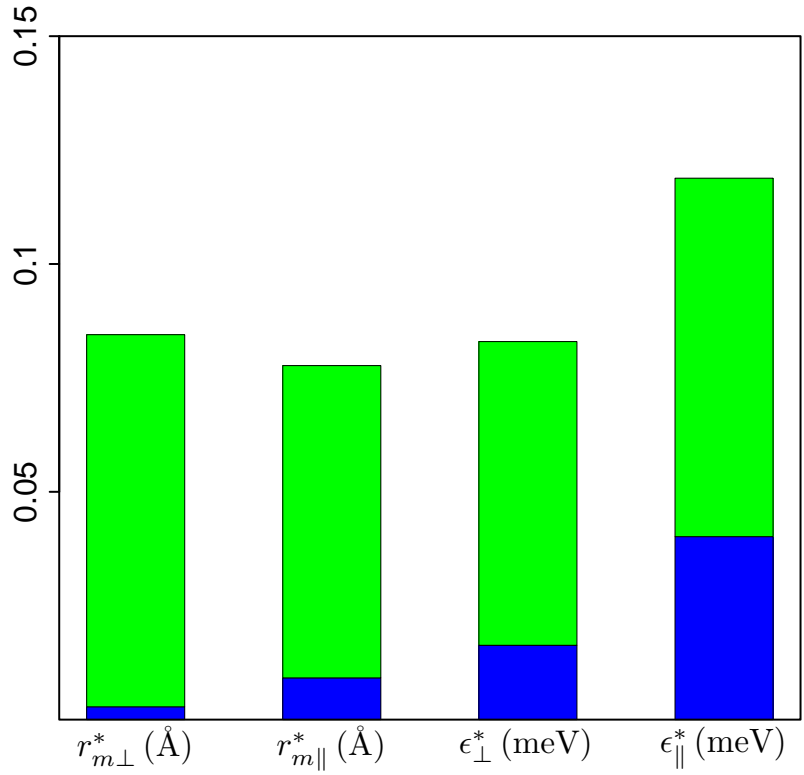}
\end{center}
\caption{The relative effect of the variation of the four PES parameters $r_{m\perp}^\ast$, $r_{m\parallel}^\ast$, $\epsilon_{m\perp}^\ast$ and $\epsilon_{m\parallel}^\ast$ on the elastic scattering cross sections. The blue area of the bars shows the uncorrelated contribution of the corresponding parameter and the green area - the joint effect that depends on the value of the other PES parameters and the collision energy.
}
\end{figure}

\begin{figure}[ht]
\label{figure7}
\begin{center}
\includegraphics[scale=0.5]{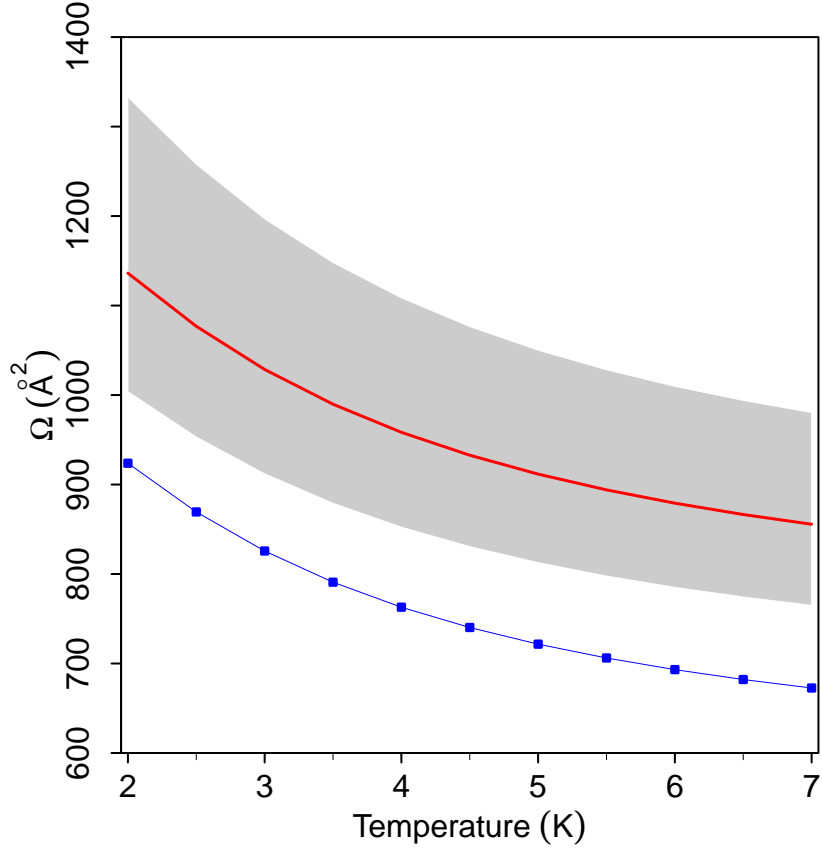}
\end{center}
\caption{The comparison of the thermally averaged cross sections for benzene - He collisions computed with the quantum CS approach (blue squares) with the 95$\%$ prediction interval of the thermally averaged cross sections for benzonitrile - He collisions (grey band). The prediction interval for the benzonitrile - He collisions and the corresponding mean values (red solid line) are obtained by integrating the results in Figure 5 with the Maxwell-Botzmann distribution and then scaling the thermally averaged cross sections by the ratio $ \tilde \Omega_{\rm qm}/ \tilde  \Omega_{\rm ct}$, as described in the text. The results are for collisions of molecules in the ground rotational state.}
\end{figure}


\clearpage
\newpage

\begin{thebibliography}{99}
\bibitem{bn-experiments}
D. Patterson and J. M. Doyle, {\it Phys. Chem. Chem. Phys.} {\bf 17}, 5372 (2015).
\bibitem{benzene-he}
Z. Li, R. V. Krems, and E. J. Heller, {\it J. Chem. Phys.} {\bf 141}, 104317 (2014).
\bibitem{benzene-he-note}
Note that Eq. (23) of Ref. \cite{benzene-he} incorrectly includes the factor 4. The correct equation is 
$\sigma_{i \rightarrow i}=\pi b_{{\rm max}}^2 {N_{i  \rightarrow i}}/{N_{\rm tot}}$. The computations reported in Ref. \cite{benzene-he} used the correct equation. 
\bibitem{scattering-theory}
J. R. Taylor, {\it Scattering Theory: the Quantum Theory of Nonrelativistic Collisions} (Dover Publications, 2006).
\bibitem{experiments-polyatomic1}
J. K\"{u}pper, F. Filsinger, and G. Meijer, {\it Faraday Discuss.} {\bf142}, 155 (2009).
\bibitem{experiments-polyatomic2}
C. Sommer, L. van Buuren, M. Motsch, S. Pohle, J. Bayerl, P. Pinkse, and G. Rempe, {\it Faraday Discuss.} {\bf 142}, 203 (2009).
\bibitem{experiments-polyatomic3}
D. Patterson, E. Tsikita, and J. M. Doyle, {\it Phys. Chem. Chem. Phys.} {\bf 12}, 9736 (2010).
\bibitem{gp-model1}
J. Sacks, S. B. Schiller, and W. J. Welch, {\it Technometrics} {\bf 31}, 41 (1989).
\bibitem{gp-model2}
T. J. Santner, B. J. Williams, and W. I. Notz, {\it The Design and Analysis of Computer Experiments} (Springer Science $\&$ Bussiness Media, New York, 2003).
\bibitem{gp-model3}
C. E. Rasmussen and C. K. I. Williams, {\it Gaussian Process for Machine Learning} (The MIT Press, Cambridge, 2006).
\bibitem{application-GP1}
D. Higdon, M. Kennedy, J. C. Cavendish, J. A. Cafeo, and R. D. Ryne, {\it SIAM J. Sci. Comput.} {\bf 26}, 448 (2004).
\bibitem{application-GP2}
D. Higdon, J. Gattiker, B. Williams, and M. Rightley, {\it J. Am. Statist. Assoc.} {\bf 103}, 570 (2008).
\bibitem{application-GP3}
R. B. Gramacy and H. K. H. Lee, {\it J. Am. Statist. Assoc.} {\bf 103}, 1119 (2008).
\bibitem{sample-size}
J. L. Loeppky, J. Sacks, and W. J. Welch, {\it Technometrics} {\bf 51}, 366 (2009).
\bibitem{previous-work}
J. Cui and R. V. Krems, accepted for publication in {\it Phys. Rev. Lett.} (arXiv:1503.01432v2).
\bibitem{ct-book}
R. B. Bernstein, {\it Atom-Molecule Collision Theory: A Guide for the Experimentalist}, (Plenum, New York, 1979).
\bibitem{adler1981geometry}
R. J. Adler, {\it The Geometry of Random Fields}, (SIAM,1981).
\bibitem{cramer2013stationary}
H. Cram\'{e}r and M. R. Leadbetter, {\it Stationary and Related Stochastic Processes: Sample Function Properties and Their Applications}, (Courier Corporation, 2013).
\bibitem{mitchell1990existence}
T. Mitchell, M. Morris, and D. Ylvisaker, {\it Stoch. Proc. Appl.} {\bf 35}, 109 (1990).
\bibitem{cressie1993statistics}
N. A. C. Cressie, {\it Statistics for Spatial Data}, (Wiley-Interscience, New York, 1993).
\bibitem{stein1999interpolation}
M. L. Stein, {\it Interpolation of Spatial Data: Some Theory for Kriging}, (Springer Science $\&$ Business Media, 1999).
\bibitem{abt1999estimating}
M. Abt, {\it Scand. J. Stat.} {\bf 26}, 563 (1999).

\bibitem{pirani2001atom}
F. Pirani, D. Cappelletti, G. Liuti, {\it Chem. Phys. Lett.}  {\bf 350},
286 (2001).

\bibitem{pirani2004atom}
F. Pirani, M. Albert\'t, A. Catro, M. Moix Teixdidor, and D. Cappelletti, {\it Chem. Phys. Lett.} {\bf 394}, 37 (2004).
\bibitem{cs-polyatomic1}
B. J. Garrison, and W. A. Lester, Jr., {\it J. Chem. Phys.} {\bf 66}, 531 (1977).
\bibitem{cs-polyatomic2}
S. Green, {\it J. Chem. Phys.} {\bf 64}, 3463 (1976).
\bibitem{private-communication}
M. Bartolomei, private communication (October 2012).
\bibitem{zhiying1}
Z. Li and E. J. Heller, {\it J. Chem. Phys.} {\bf 136}, 054306 (2012).
\bibitem{jie-zhiying}
J. Cui, Z. Li, and R. V. Krems, {\it J. Chem. Phys.} {\bf 141}, 164315 (2014).
\bibitem{saltelli2009sensitivity}
A. Saltelli, K. Chan, and E. M. Scott, {\it Sensitivity Analysis} (Wiley, New York, 2009).
\bibitem{saltelli2008global}
A. Saltelli, M. Ratto, T. Andres, F. Campolongo, J. Cariboni, D. Gatelli, M. Saisana, and S. Tarantola, {\it Global Sensitivity Analysis: the Primer} (John Wiley $\&$ Sons, 2008).
\bibitem{roustant2012dicekriging}
O. Roustant, D. Ginsbourger, and Y. Deville, {\it J. Stat. Softw.} {\bf 51}, 1 (2012).


\end{thebibliography}
\end{document}